\documentclass[10pt,journal,a4paper,oneside,twocolumn]{IEEEtran}

\makeatletter
\def\@endtheorem{\endtrivlist}
\makeatother

\usepackage{geometry}
\usepackage{graphicx}
\usepackage{amsmath}
\newtheorem{Definition}{Definition}
\newtheorem{Lemma}{Lemma}

\newcommand{\BibTeX}{{\rmfamily B\kern-.05em{\scshape i\kern-.025em
b}\kern-.08em T\kern-.1667em\lower.7ex\hbox{E}\kern-.125emX}}

\begin{document}

\title{
%Calculating Its Position\\-
A Sorting Algorithm Based on Calculation}

\author{\authorblockN{Sheng~Bao} ~\IEEEmembership{Student Member,~IEEE,} \\ \authorblockA{Dept. of Information Engineering,Nanjing Univ. \ of P \& T,Nanjing 210046,CHINA \\ Email : forrest.bao@gmail.com} \\\authorblockN{De-Shun~Zheng}\\ \authorblockA{Dept. of Telecommunication Engineering,Nanjing Univ of P \& T,Nanjing 210046,CHINA \\ Email : gtzds@163.com}}

\maketitle

% \begin{flushright}
% \textit{"Come along,children.Follow me."\\
% Before you could wink an eyelash\\
% Jack,Knak,Lack,Mack,Nack,Ouack,Pack, and Quack\\
% fell into line,just as they had been taught.}\\
% ---Robert McCloskey
% \end{flushright}

\begin{abstract}
This article introduces an adaptive sorting algorithm that can relocate elements accurately by substituting their values into a function. %which we name it the guessing function. %whose time complexity is $O(n)$.
%The essential of sorting is relocating elements by their values.
We focus on building this function which is the mapping relationship between record values and their corresponding sorted locations essentially.
The time complexity of this algorithm $O(n)$,when records distributed uniformly.
Additionally, similar approach can be used in the searching algorithm.

\end{abstract}


\IEEEpeerreviewmaketitle

\section{Introduction}

%Sorting are the problem that devotes to locate elements according to their values. Sorting algorithms identify the new location of every element in the ordered array and then rearrange them based on their perspective new postition.

We live in a world obsessed with keeping information, and to find it,we must keep it in some sensible order.\cite{Data_Structures_and_Program_Design_in_C++}
Computers spend a considerable amount of their time keeping data in order.\cite{JAVA}
The objective of the sorting method is to rearrange the records so that their keys are ordered according to some well-defined ordering rules.\cite{Algorithms_in_C++}

The essense of sorting is a mapping relationship between record values and their corresponding ordered positions. A perfect sorting algorithm will make us accomplish our goal via just one calculation,substituting the value of elements into the function and returning us their location.

This article describes a new sorting algorithm which devotes to implement the mapping relationship mentioned previously.
Assuming the mapping ralationship is linear,we devised two approaches.One depends on the maximum and the minimum value of
records, the other depends on the statistic property of records.Of course,the second one takes more time in determining the mapping relationship.

To make the mapping more accurate,the second pass mapping on the intervals where records density is high are devised.

This algorithm consists of two parts,mapping routine and post-mapping routine.They will both be discussed.

Performance of this algorithm are also discussed.In the condition of uniform distribution,the time complexity is $O(n)$.

\section{Preliminaries}
In following sections, we will describe our algorithm of sorting an array of elements which we call records.
All array positions contain out-of-order records that are assumed to be sorted.
To simplify matters, we assume these records are all real numbers of the type double. \footnote{This ``double" type is defined by ANSI C++ standard.} And more, we assume that all of our operations can be done in main memory.

In following discussion, the number of records is denoted as $N$.
The routine of sorting is considered as putting $N$ records into prepared $N$ boxes. To identify these boxes,they are assigned  indics which are integers in interval $[1,N]$.After the sorting routine, records should locate in boxes ascendly.

At the end of this section,we name the function that will be introduced as guessing function.It is named from one of its properties
is ``guessing" the location of records.The routine of substituting records into guessing function is called mapping.
\section{Building the guessing function}
\subsection{Basic properties of guessing function}
The guessing function is defined following.
\begin{Definition}
Guessing function is such a function whose argument is the value of a record and returning value is the location of this record after soring.
\end{Definition}
The ideal guessing function should have following properties :

\begin{itemize}
\item It should be a single function.
\item The function range should be values of the maximum and the minimum records.
\item The function domain should be $[1,N]$.
\end{itemize}

It is easily to infer that the minimum record should be put into the first box whereas the maximum record should be put into the last box.Denote the maximum value of elements as $X_{max}$ while the minimum value of elements as $X_{min}$.
%$x_{max}$ and $x_{min}$ represent elements whose value are $X_{max}$ and $X_{min}$ respectively.

\subsection{two terminals  approach}

Based on the idea of building the function as simple as possible,we assume guessing function as a linear function with two ternimals ,$(X_{min},1)$ and $(X_{max},N)$.The equation of guessing function is
\begin{equation}
\frac {x-X_{min}}{X_{max}-X_{min}}= \frac {n-1} {N-1}
\end{equation}
where $x$ is the value of a record and $n$ is the box index where the record locates.

Thus
\begin{equation}
n=\frac {x-X_{min}}{X_{max}-X_{min}} (N-1)+1
\end{equation}

Since the indics of boxes are integers, so we need to round $n$ down.Then we obtain the simplest guessing function.
\begin{equation}
g_{1}(x)=\left \lfloor {\frac {x-X_{min}} {X_{max}-X_{min}} (N-1)}
\right \rfloor +1
\end{equation}

%This function guarantees that larger elements will be mapped to positions with larger indics whereas smaller elements will be mapped to positions with smaller indics.

%Denote the tangent of guessing function as $k$ where $k=\frac {N-1} {X_{max}-X_{min}}$.

\begin{Definition}
Global tangent is defined as the tangent of guessing function of all the records.
\begin{equation}
k_{global}={N-1 \over X_{max}-X_{min}}
\label{global tangent}
\end{equation}
\end{Definition}
The reason why we call it ``global tangent" will be explained later.

Guessing function can be rewrited as
\begin{equation}
g_{1}(x)=\left \lfloor (x-X_{min}) k_{global}
\right \rfloor +1
\label{GFI}
\end{equation}

\subsection{An alternative approach}
We also devised an alternative approach that has general adaption to normally distributed record.

According to the property of Guassian distribution, almost the entire elements lie in the symmetric interval $(M-3\sigma,M+3\sigma)$,
where M is the mean and $\sigma$ is the standard deviation.\cite{Statistics}

We can assume the difference between record's value and mean lies in the interval $(-3\sigma,3\sigma)$ while their corresponding
box indics lies from 1 to N.
%The difference between element's location and center in sorted sequence if proportional to the difference between element's value and mean. So we can also define
So we can also define $k_{global}$ as ${n \over 6\sigma}$.
But there is a difference compared with the first approach.Such mapping may lead to box index greater than $N$ or less than 1.So a round routine is needed to limit box index in $[1,N]$.

Since this approach needs at least two passes to obtain statistic information and it need to judge every index,it will elapse much time than the first one in building guessing function.But it's mapping may be more accurate.

\subsection{Hash table and guessing function}
Some one will consider our method is just like a hash table.But in fact they are based on different principles.And more,the guessing function can be extended to a more precisely one.

\subsection{More precisely mapping:guessing function II}
No matter which approach is adopted,one disadvantage of previous defined guessing function is that records with similiar values will be mapped into same boxes.This is because the tangent of guessing function that we used is a contant.
An improved function that uses variable tangent can map elements more accurately since its tangent is adaptive to the density of record values.
For we are going to introduce a better function, we denote the function in eq.\ref{GFI} as Guessing Function $\mathrm{I}$ and the following function as Guessing Function $\mathrm{II}$.

Guessing function II is based on guessing function $\mathrm{I}$.The only difference is the tangent of Guessing Function II is a variable.The routine on every box is the same as the one that performs in guessing function $\mathrm{I}$.The distribution array,whose element is denoted as $A[n]$,should be defined here.
\begin{Definition}
Distribution array is such an array that its scale equals to $N$ whereas the value of $A[n]$ is the sum of record numbers in boxes whose indics are not greater than $n$.
\end{Definition}

Of course,the value of array element whose index is less than 1 or greater than N is $0$.

Then we can infer that
\begin{Lemma}
the final position of elements in the $n$th box is between $A[n-1]+1$ and $A[n]$.
\label{position range}
\end{Lemma}

To any record,we have
\begin{Lemma}
\begin{equation}
\frac {n-1} {k_{global}} + X_{min} \leq x < \frac {n} {k_{global}} +X_{min}
\end{equation}
where x is its value and n is the index of the box where it is mapped by guessing function I.
\label{box value}
\end{Lemma}

%According to Lemma \ref{position range}, the final position of this element is between $A[n-1]$ and $A[n]$.
%Then we will apply guessing function $\mathrm{I}$ onto this box.
Combining Lemma \ref{position range} and Lemma \ref{box value},we obtain that the guessing function $\mathrm{I}$ in this box has two terminals s,$(\frac {n-1}{k_{global}} +X_{min},A[n-1]+1)$ and $(\frac {n}{k_{global}} +X_{min},A[n])$.

Specially,if this box is the first box,where $n=1$,terminals of guessing function will be $(X_{min},1)$ and $(\frac {1}{k_{global}} +X_{min},A[1])$.
In the last the box,the terminals should be $(\frac {N-1} {k_{global}+X_{min}} ,A[N-1]+1)$ and $({N \over k_{global}+X_{min}},A[N])$.
%Attention we don't consider guessing function in the last box,since there is no box whose index is greater than it.

Then we can consider each box independently.Before we applying guessing function I onto each box,the local
tangent of guessing function should be introduced.

\begin{Definition}
Local tangent of guessing function is defined as the tangent of the line that passes through point
$(\frac {n-1}{k_{global}} +X_{min},A[n-1]+1)$ and $(\frac {n}{k_{global}} +X_{min},A[n])$
\end{Definition}

This definition is the reason why we call the tangent defined in eq.\ref{global tangent} as global tangent.

So we have
\begin{eqnarray}
k_{local}&=&\frac
{A[n]-A[n-1]-1}
{\Big ( {N \over k_{global}}+X_{min}\Big )-
\Big ( {N-1 \over k_{global}}+X_{min} \Big )}\nonumber\\
&=&k_{global}(A[n]-A[n-1]-1)
\label{klocal}
\end{eqnarray}

Substituting above information into eq.\ref{GFI},we obtain the local guessing function in a box as
\begin{equation}
\left \lfloor \Big [ x-\frac {n-1}{k_global}+X_{min} \Big]k_{local} \right \rfloor+1
\end{equation}

% Thus,
% \begin{eqnarray}
% \left \lfloor \Big [ x-\frac {n-1}{k_{global}}+X_{min} \Big]k_{global}(A[n]-A[n-1]-1) \right \rfloor
% +1
% \end{eqnarray}

Considering position of elements in this box starts from $A[n]$ , we obtain the guessing function of the entire records

\begin{equation}
g_{2}(x)=
% \begin{cases}
% A[n]+\left \lfloor \Big [ x-\frac {n-1}{k_{global}}+X_{min} \Big]k_{local} \right \rfloor +1 , 1 < n \leq N \\
% A[1]+\left \lfloor \Big [ x+X_{min} \Big]k_{global}(A[n]-1) \right \rfloor+1 , n=1
% \end{cases}
A[n]+\left \lfloor \Big [ x-\frac {n-1}{k_{global}}+X_{min} \Big]k_{local} \right \rfloor +1
\label{GFII}
\end{equation}

where $n$ is calculated by eq.\ref{GFI} and $k_{local}$ is given by eq.\ref{klocal}.

We name eq.\ref{GFII} as Guessing Function $\mathrm{II}$.

\subsection{The neccessarity of guessing function $\mathrm{II}$}
Some of our test indicate that the time elapsed by guessing function $\mathrm{II}$ is almost 5 times than the one of guessing funciton
$\mathrm{I}$.If your record distribution is similar to uniform distribution,guessing function $\mathrm{I}$ is enough.But if your record distribution is gathered in some intervals,maybe guessing function $\mathrm{II}$ is needed.

\section{Post-mapping routines}
No matter guessing function $\mathrm{I}$ or guessing function $\mathrm{II}$,we can't guarantee that every box contains only one record.
To records in a same box,we apply traditional sorting algorithms to sort them so that each box is sorted.One pass travesal will retrieve them out and return us a sorted array.

\section{Performance analysis}
\subsection{Time complexity}
In uniform distribution condition,the time compelxity of our algorithm is $O(n)$.

\begin{proof}
The probability of an element being mapped into any box is $1/N$ equally.We can infer the probability of a box contains no element is
${0 \choose N} (1-{1 \over {N}})^{N}=(1-{1 \over N})^{N}$
And we have
$$
\lim_{N \to \infty} {(1-{1 \over N})^{N} = e^{-1}}
$$

So the expectation of boxes which contain no element is $e^{-1}N$.

After the first pass mapping,N elements are mapped into $(1-e^{-1})N$ boxes.In these boxes,the expectation of element amount in these boxes is $\frac {1} {1-e^{-1}}$ per box. Considering the final position of every element should be limited in the box where it is mapped into,in the second time of mapping,the expectation of error interval of mapping is less than $\frac {1} {2(1-e^{-1})}$.

So N elements need totally less than $\frac {N} {2(1-e^{-1})}$ times of move.Considering the mapping operation and the operation of constructing array A[n] have linear time complexity,we can conclude the time complexity is $O(n)$.\cite{Computer_Algorithms:_Introduction_to_Design_and_Analysis}
\end{proof}

\subsection{Space complexity}
Before mapping,the space for storing result of guessing function $\mathrm{I}$ and $\mathrm{II}$is proportional to $N$.
Also the space for distribution array is proportional to $N$.
Space for storing other variable is constant.So the space complexity of both guessing function $\mathrm{I}$ and $\mathrm{II}$ are both $O(n)$.

\section{Comparation with other sorting algorithms}
\begin{figure*}[!htb]
\centering \caption[fig]{time vs. scale to uniform distributed
records}
\includegraphics[scale=0.7]{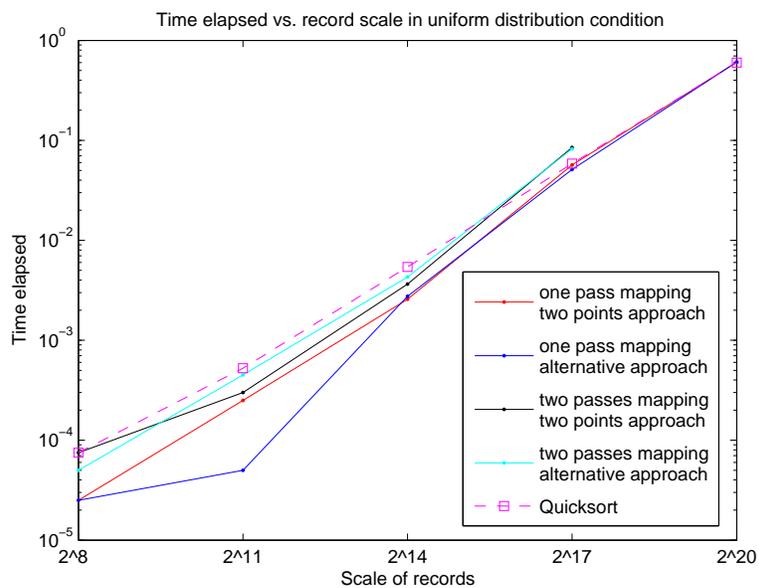}
\label{uniformfig}
\end{figure*}

% Some tests are performed whose results are given in
% Fig. \ref{uniformfig} and Fig. \ref{normalfig}
Some tests are performed on a computer whose CPU is AMD Athlon
2000+ and OS is Fedora Core 1(Linux Kernel 2.4.22-1). Testing
programmes are executed at multiuser text mode while compiled by
gcc 3.3.2 without optimization.
Uniformly distributed numbers ranging from $-20000000$ to $20000000$ are generated and are sorted in testing programmes.
Table \ref{comtable} lists the sorting time of different algorithms when the scale of record increases.
Fig.\ref{uniformfig} also illustrates the time elpased comparison with some other algorithms.
\begin{table*}
\caption[comparison]{Sorting time of different algorithms}
\begin{center}
\begin{tabular}[!ht]{c | r@{.}l r@{.}l r@{.}l r@{.}l r@{.}l}
\hline  Algorithms \slash Scale     & \multicolumn{2}{c}{$2^8$} & \multicolumn{2}{c}{$2^{11}$}  &\multicolumn{2}{c}{$2^{14}$}  &\multicolumn{2}{c}{$2^{17}$}  &\multicolumn{2}{c}{$2^{20}$}\\ \hline
%\hline Bubble Sort\cite{Algorithms_in_C++} & 0&0007        &0&04495    & 2&972725  & N&A.      & N&A.\\
%\hline Insertion Sort\cite{Algorithms_in_C++}  & 0&0002461538462   &0&0156     & 1&005369231   & 153&22    & N&A.\\
%\hline Mergesort\cite{Algorithms_in_C++}   & 0&00009230769231  &0&001      & 0&01647692308 & N&A.      & N&A.\\
\hline  Quicksort\cite{Algorithms_in_C++}   & 0&000075  &0&000525& 0&005425 & 0&058475  & 0&600225\\
%\hline Radix Exchange Sort\cite{Algorithms_in_C++} &0&0001076923077    &0&001169230769 &0&008369230769 & 0&08425   & 0&7815\\
%\hline  Selection Sort\cite{Algorithms_in_C++} & 0&0003692307692   & N&A.      & N&A.      & N&A.      & N&A.\\
%\hline Shellsort\cite{Data_structure_and_algorithm_analysis}   & 0&0001076923077   &0&001723076923 &0&02475384615  &0&529      & 11&361\\
\hline  Guessing function
   \\one pass mapping
   \\two points approach
           & 0&000025  &0&00025 &0&002575  &0&056725   &0&603525\\
\hline  Guessing function
   \\one pass mapping
   \\alternative approach
           & 0&000025  &0&00005 &0&00275   &0&05105    &0&60855\\
\hline  Guessing function
   \\two passes mapping
   \\two points approach
           & 0&000075  &0&0003 &0&00365    &0&0848.        &N&A.\\
\hline  Guessing function
   \\two passes mapping
   \\alternative approach
           & 0&00005   &0&00045 &0&0043    &0&081975       &N&A.\\
\hline
\end{tabular}
\label{comtable}
\end{center}
\end{table*}

%%%%%%%%%%%%%%%%%%%%%%%%%%%%%%%%%%%%%%%%%%%%%%%%%%%%%%%%%%%%%%%%
%Deshun Zheng's experiment
% \begin{figure}[!htb]
% %\centering
% \includegraphics[scale=0.4]{uniform_zhen.eps}
% \caption[fig]{time vs. scale to uniform
% distributed records}
% \label{uniformfig}
% \end{figure}
%
% \begin{figure}[!htb]
% \centering
% \includegraphics[scale=0.4]{normal_zhen.eps}
% \caption[fig]{time vs. scale to normal
% distributed records}
% \label{normalfig}
% \end{figure}
%
% Fig. \ref{uniformfig} and Fig. \ref{normalfig} illustrate the time
% elapsed vs. records scale of our algorithms and Quicksort. It is
% obvious that in small and medium scale,our algorithm performs
% better than Quicksort.
%
% Respectively,Fig. \ref{uniformfig} indicates that guessing
% functions are linear approximately in uniform distribution
% condition while in normal distribution the time elapsed by
% Quicksort reduces and the ones of guessing function 1 and guessing
% function 2 increases.

%These tests also indicate the guessing function is very important in our algorithm.Once we used a code that the mean of record is calculated incorrectly.The time elapsed became very long.

\section*{Acknowledgement}
Mr. Michael Shell in Dept. of Electrical and Computer Engineering of Georgia Institute of Technology,who is also the author of IEEETran
\LaTeX  \ class and \BibTeX\ style package,gave us lots of instructions about using those two packages in our typesetting process.

\bibliographystyle{IEEEtran}
\bibliography{guessingbib}

\end{document}